\documentclass{article}      

\usepackage{epsfig}

\newcommand{\ba}{\begin{eqnarray}}
\newcommand{\ea}{\end{eqnarray}}
\begin{document}

\title{Mass spectrum of pentaquarks} 

\author{R. Bijker\\
Instituto de Ciencias Nucleares, 
Universidad Nacional Aut\'onoma de M\'exico,\\
A.P. 70-543, 04510 M\'exico, D.F., M\'exico
\and
M.M. Giannini and E. Santopinto \\
Dipartimento di Fisica dell'Universit\`a di Genova, 
I.N.F.N., Sezione di Genova, \\
via Dodecaneso 33, 16164 Genova, Italy}
\date{August, 2004}
\maketitle                   

\begin{abstract}
We discuss the properties of the pentaquark in a collective stringlike 
model with a nonplanar configuration of the four quarks and the antiquark.  
In an application to the mass spectrum of exotic $\Theta$ baryons, 
we find that the ground state pentaquark has angular momentum and parity 
$J^p=1/2^-$ and a small magnetic moment of 0.382 $\mu_N$. The decay width 
is suppressed by the spatial overlap with the decay products. 
\end{abstract}

\maketitle

\section{Introduction}

The building blocks of atomic nuclei, the nucleons, are composite extended 
objects, as is evident from the large anomalous magnetic moment, the 
excitation spectrum and the charge distribution of both the proton and 
the neutron. To first approximation, the internal structure of the nucleon 
at low energy can be ascribed to three bound constituent quarks $q^3$. 
The nucleon and its excited states, collectively known as baryons,  
are accommodated into flavor singlets, octets and decuplets. 
The strangeness of the known baryons is either zero (nucleon, $\Delta$) 
or negative ($\Lambda$, $\Sigma$, $\Xi$ and $\Omega$). Baryons with quantum 
numbers that cannot be obtained from triplets of quarks are called exotic. 

Until recently, there was no experimental evidence for the existence of such 
exotic baryons. The discovery of the $\Theta(1540)$ baryon with positive 
strangeness $S=+1$ by the LEPS Collaboration \cite{leps} as the 
first example of an exotic baryon, has motivated an enormous amount of 
experimental and theoretical studies \cite{workshop}. 
The width of this state is observed to be very small 
$< 20$ MeV (or perhaps as small as a few MeV's). More recently, the NA49 
Collaboration \cite{cern} reported evidence for the existence of another 
exotic baryon $\Xi(1862)$ with strangeness $S=-2$. The $\Theta^+$ and 
$\Xi^{--}$ resonances are interpreted as $q^4 \bar{q}$ pentaquarks belonging 
to a flavor antidecuplet with quark structure $uudd\bar{s}$ and $ddss\bar{u}$, 
respectively. In addition, there is now the first evidence \cite{h1} for a 
heavy pentaquark $\Theta_c(3099)$ in which the antistrange quark in the 
$\Theta^+$ is replaced by an anticharm quark. The spin and parity of these 
states have not yet been determined experimentally. 
For a review of the experimental status we refer to \cite{zhaoclose}. 

Theoretical interpretations range from chiral soliton models \cite{soliton} 
which provided the motivation for the experimental searches, correlated 
quark (or cluster) models \cite{cluster}, and various constituent quark 
models \cite{cqm,BGS}. A review of the theoretical literature of 
pentaquark models can be found in \cite{jennings}. 

In this contribution, we discuss a collective stringlike model of 
$q^4\bar{q}$ pentaquarks in which the four quarks are located at the 
corners of an equilateral tetrahedron and the antiquark in its center. 
This nonplanar configuration arises as a consequence of the permutation 
symmetry of the four quarks \cite{BGS1}. As an application, we discuss the 
spectrum of exotic $\Theta$ baryons, as well as the parity and magnetic 
moments of the ground state decuplet baryons. The decay width is suppressed 
by the spatial overlap with the decay products. 

\section{Pentaquark states}

We consider pentaquarks to be built of five constituent parts whose 
dynamics is characterized by both internal and spatial degrees of freedom. 

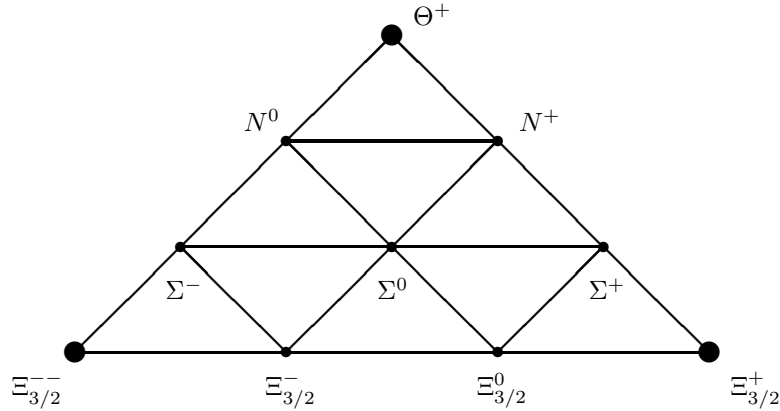
\begin{figure}
\centering
\setlength{\unitlength}{0.8pt}
\begin{picture}(350,185)(70,80)
\thicklines
\put(200,200) {\line(1,0){100}}
\put(150,150) {\line(1,0){200}}
\put(100,100) {\line(1,0){300}}
\put(100,100) {\line(1,1){150}}
\put(200,100) {\line(1,1){100}}
\put(300,100) {\line(1,1){ 50}}
\put(200,100) {\line(-1,1){ 50}}
\put(300,100) {\line(-1,1){100}}
\put(400,100) {\line(-1,1){150}}
\multiput(250,250)(100,0){1}{\circle*{5}}
\multiput(200,200)(100,0){2}{\circle*{5}}
\multiput(150,150)(100,0){3}{\circle*{5}}
\multiput(100,100)(100,0){4}{\circle*{5}}
\multiput(100,100)(300,0){2}{\circle*{10}}
\put(250,250){\circle*{10}}
\put(260,255){$\Theta^+$}
\put(180,205){$N^0$}
\put(310,205){$N^+$}
\put(143,125){$\Sigma^-$}
\put(243,125){$\Sigma^0$}
\put(343,125){$\Sigma^+$}
\put( 70, 80){$\Xi_{3/2}^{--}$}
\put(190, 80){$\Xi_{3/2}^{-}$}
\put(290, 80){$\Xi_{3/2}^{0}$}
\put(410, 80){$\Xi_{3/2}^+$}
\end{picture}
\caption{\small $SU(3)$ antidecuplet. The isospin-hypercharge multiplets are 
$(I,Y)=(0,2)$, $(\frac{1}{2},1)$, $(1,0)$ and $(\frac{3}{2},-1)$. 
Exotic states are located at the corners and are indicated with $\bullet$.}
\label{flavor33e}
\end{figure}

The internal degrees of freedom are: a flavor triplet for the quarks and a 
flavor anti-triplet for the antiquark (for the three light flavors: up, 
down and strange), a spin doublet (for spin $s=1/2$) and a color triplet. 
The corresponding algebraic structure consists of the usual spin-flavor 
and color algebras $SU_{\rm sf}(6) \otimes SU_{\rm c}(3)$. The full 
decomposition of the spin-flavor states into spin and flavor states can 
be found in \cite{BGS}
\ba
SU_{\rm sf}(6) \supset SU_{\rm f}(3) \otimes SU_{\rm s}(2) 
\supset SU_{\rm I}(2) \otimes U_{\rm Y}(1) \otimes SU_{\rm s}(2) ~.
\label{sfchain}
\ea
The allowed flavor multiplets are singlets, octets, decuplets, antidecuplets,  
27-plets and 35-plets. The first three have the same values of the isospin 
$I$ and hypercharge $Y$ as $q^3$ systems. However, the antidecuplets, 
the 27-plets and 35-plets contain exotic states which cannot be obtained 
from three-quark configurations. The latter states are more easily identified 
experimentally due to the uniqueness of their quantum numbers. 
The recently observed $\Theta^+$ and $\Xi^{--}$ resonances are 
interpreted as pentaquarks belonging to a flavor antidecuplet with 
isospin $I=0$ and $I=3/2$, respectively. In Fig.~\ref{flavor33e} 
the exotic states are indicated by a $\bullet$: the $\Theta^+$ is an 
isosinglet $I=0$ with hypercharge $Y=2$ (strangeness $S=1$), and the 
cascade pentaquarks $\Xi_{3/2}$ have hypercharge $Y=-1$ (strangeness $S=-2$) 
and isospin $I=3/2$. 

In the construction of the classification scheme we are guided by two 
conditions: the pentaquark wave function should be a color singlet and it 
should be antisymmetric under any permutation of the four quarks \cite{BGS}. 
The permutation symmetry of the four-quark system is given by the 
permuation group $S_4$ 
which is isomorphic to the tetrahedral group ${\cal T}_d$. We use 
the labels of the latter to classify the states by their symmetry 
character: symmetric $A_1$, antisymmetric $A_2$ or mixed symmetric 
$E$, $F_2$ or $F_1$, corresponding to the Young tableaux $[4]$, $[1111]$, 
$[22]$, $[211]$ and $[31]$, respectively. 

The relative motion of the five constituent parts is described in terms 
of the Jacobi coordinates 
\ba
\vec{\rho}_1 &=& \frac{1}{\sqrt{2}} ( \vec{r}_1 - \vec{r}_2 ) ~, 
\nonumber\\
\vec{\rho}_2 &=& \frac{1}{\sqrt{6}} ( \vec{r}_1 + \vec{r}_2 - 2\vec{r}_3 ) ~, 
\nonumber\\
\vec{\rho}_3 &=& \frac{1}{\sqrt{12}} 
( \vec{r}_1 + \vec{r}_2 + \vec{r}_3 - 3\vec{r}_4 ) ~, 
\nonumber\\
\vec{\rho}_4 &=& \frac{1}{\sqrt{20}} 
( \vec{r}_1 + \vec{r}_2 + \vec{r}_3 + \vec{r}_4 - 4\vec{r}_5 ) ~, 
\label{jacobi}
\ea
where $\vec{r}_i$ ($i=1,..,4$) denote the coordinate of the $i$-th quark,  
and $\vec{r}_5$ that of the antiquark. The last Jacobi coordinate 
is symmetric under the interchange of the quark coordinates, 
and hence transforms as $A_1$ under ${\cal T}_d$, whereas 
the first three transform as three components of $F_2$.  

The total pentaquark wave function is the product of the spin, flavor, 
color and orbital wave functions. 
Since the color part of the pentaquark wave function is a singlet 
and that of the antiquark an anti-triplet, the color wave function of 
the four-quark configuration is a triplet with $F_1$ symmetry.  
The total $q^4$ wave function is antisymmetric ($A_2$), hence the 
orbital-spin-flavor part has to have $F_2$ symmetry 
\ba
\psi \;=\; \left[ \psi^{\rm c}_{F_1} \times 
\psi^{\rm osf}_{F_2} \right]_{A_2} ~.
\label{wf}
\ea
Here the square brackets $[\cdots]$ denote the tensor coupling under the 
tetrahedral group ${\cal T}_d$. 

\section{Stringlike model}

\begin{figure}
\hspace{-1cm} \centerline{\hbox{
\epsfig{figure=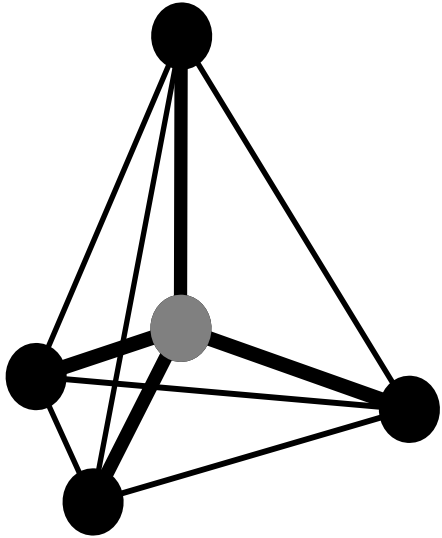,height=0.5\textwidth,width=0.5\textwidth} }}
\caption[]{\small Geometry of stringlike pentaquarks}
\label{tetrahedron}
\end{figure}

In this section, we discuss a stringlike model for pentaquarks, which is a 
generalization of a collective stringlike model developed for $q^3$ baryons 
\cite{BIL}. The general approach is that of introducing an interacting 
boson model to describe the orbital excitations of the pentaquark. 
We introduce a dipole boson with $L^p=1^-$ for each independent relative 
coordinate, and an auxiliary scalar boson with $L^p=0^+$, which leads to a 
compact spectrum-generating algebra of $U(13)$ for the radial excitations. 
As a consequence of the invariance of the interations under the permutation  
symmetry of the four quarks, the most favorable geometric configuration is an 
equilateral tetrahedron in which the four quarks are located at the four 
corners and the antiquark in its center \cite{BGS1} 
(see Fig.~\ref{tetrahedron}). 
This configuration was also considered in \cite{song}  
in which arguments based on the flux-tube model were used to suggest a 
nonplanar structure for the $\Theta(1540)$ pentaquark to explain its narrow 
width. In the flux-tube model, the strong color field between a pair of a 
quark and an antiquark forms a flux tube which confines them. 
For the pentaquark there would be four such flux tubes connecting the  
quarks with the antiquark. 

\subsection{Mass spectrum of $\Theta$ baryons}

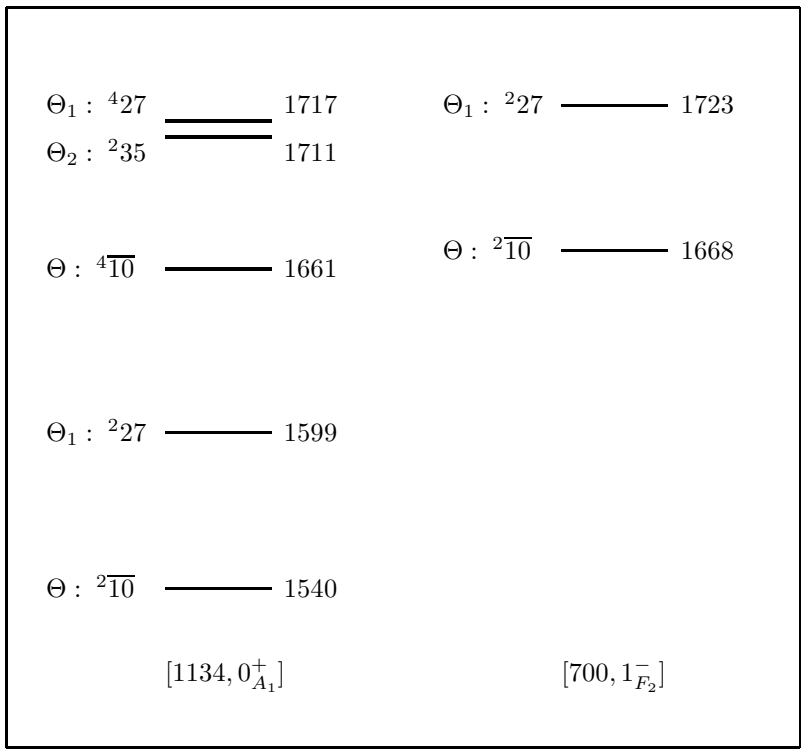
\begin{figure}
\centering
\setlength{\unitlength}{1.0pt}
\begin{picture}(300,280)(0,0)
\put(  0,  0) {\line(1,0){300}}
\put(  0,280) {\line(1,0){300}}
\put(  0,  0) {\line(0,1){280}}
\put(300,  0) {\line(0,1){280}}

\thicklines
\put( 60, 60) {\line(1,0){ 40}}
\put( 60,181) {\line(1,0){ 40}}
\put( 60,119) {\line(1,0){ 40}}
\put( 60,237) {\line(1,0){ 40}}
\put( 60,231) {\line(1,0){ 40}}
%\put( 50, 40) {$[42111]_{F_2}$}
\put( 60, 25) {$[1134,0^+_{A_1}]$}

\put(210,188) {\line(1,0){ 40}}
\put(210,243) {\line(1,0){ 40}}
%\put(200, 40) {$[51111]_{A_1}$}
\put(210, 25) {$[700,1^-_{F_2}]$}

\put( 15, 57) {$\Theta: \; ^{2}\overline{10}$}
\put( 15,178) {$\Theta: \; ^{4}\overline{10}$}
\put( 15,116) {$\Theta_1: \; ^{2}27$}
\put( 15,240) {$\Theta_1: \; ^{4}27$}
\put( 15,222) {$\Theta_2: \; ^{2}35$}

\put(165,185) {$\Theta: \; ^{2}\overline{10}$}
\put(165,240) {$\Theta_1: \; ^{2}27$}

\put(105, 57) {1540}
\put(105,178) {1661}
\put(105,116) {1599}
\put(105,240) {1717}
\put(105,222) {1711}

\put(255,185) {1668}
\put(255,240) {1723}
\end{picture}
\vspace{15pt}
\caption{\small Spectrum of $\Theta$ pentaquarks. Masses are given in MeV.}
\label{theta}
\end{figure}

Hadronic spectra are characterized by the occurrence of linear Regge 
trajectories with almost identical slopes for baryons and mesons. 
Such a behavior is also expected on basis of soft QCD strings in which the 
strings elongate as they rotate. In the same spirit as for algebraic models 
of stringlike $q^3$ baryons \cite{BIL}, we use the mass-squared operator 
\ba
M^2 \;=\; M_0^2 + M_{\rm vib}^2 + M_{\rm rot}^2 + M_{\rm sf}^2 ~. 
\label{mass}
\ea
The vibrational term $M_{\rm vib}^2$ describes the vibrational spectrum 
corresponding to the normal modes of a tetrahedral $q^4 \overline{q}$ 
configuration 
\ba
M_{\rm vib}^2 \;=\; \epsilon_1 \, \nu_1 
+ \epsilon_2 \, \left( \nu_{2a}+\nu_{2b} \right)  
+ \epsilon_3 \, \left( \nu_{3a}+\nu_{3b}+\nu_{3c} \right)  
+ \epsilon_4 \, \left( \nu_{4a}+\nu_{4b}+\nu_{4c} \right) ~. 
\label{mvib}
\ea
The rotational energies are given by a term linear in the orbital angular 
momentum $L$ which is responsable for the linear Regge trajectories in 
baryon and meson spectra 
\ba
M^2_{\rm rot} \;=\; \alpha \, L ~.
\label{mrot}
\ea
The spin-flavor part is expressed in a G\"ursey-Radicati form, 
i.e. in terms of Casimir invariants of the spin-flavor groups of 
Eq.~(\ref{sfchain})
\ba
M^2_{\rm sf} \;=\; a \, C_{2SU_{\rm sf}(6)} + b \, C_{2SU_{\rm f}(3)} 
+ c \, C_{2SU_{\rm s}(2)} + d \, C_{1U_{\rm Y}(1)} 
+ e \, C_{1U_{\rm Y}(1)}^2 + f \, C_{2SU_{\rm I}(2)} ~.
\label{msf}
\ea
The coefficients $\alpha$, $a$, $b$, $c$, $d$, $e$ and $f$ are taken 
from a previous study of the nonstrange and strange baryon resonances 
\cite{BIL}, and the constant $M_0^2$ is determined by identifying the 
ground state exotic pentaquark with the recently observed 
$\Theta(1540)$ resonance. Since the lowest orbital states with $L^p=0^+$ 
and $1^-$ are interpreted as rotational states, for these excitations there 
is no contribution from the vibrational terms $\epsilon_1$, $\epsilon_2$, 
$\epsilon_3$ and $\epsilon_4$. The results for the lowest $\Theta$ pentaquarks 
(with strangeness $S=+1$) are shown in Fig.~\ref{theta}.

The lowest pentaquark belongs to the flavor antidecuplet with spin $s=1/2$ 
and isospin $I=0$, in agreement with the available experimental information 
which indicates that the $\Theta(1540)$ is an isosinglet. In the present 
calculation, the ground state pentaquark belongs to the $[42111]_{F_2}$ 
spin-flavor multiplet, indicated in Fig.~\ref{theta} by its dimension 1134, 
and an orbital excitation $0^+$ with $A_1$ symmetry. Therefore, the ground 
state has angular momentum and parity $J^p=1/2^-$, in agreement with recent 
work on QCD sum rules \cite{sumrule} and lattice QCD \cite{lattice}, 
but contrary to the chiral soliton model \cite{soliton}, various cluster 
models \cite{cluster} and a lattice calculation \cite{chiu} that predict a 
ground state with positive parity. The first excited state at 1599 MeV is 
an isospin triplet $\Theta_1$-state of the 27-plet 
with the same value of angular momentum and parity $J^p=1/2^-$. The lowest 
pentaquark state with positive parity occurs at 1668 MeV and belongs to the 
$[51111]_{A_1}$ spin-flavor multiplet (with dimension 700) and an orbital 
excitation $1^-$ with $F_2$ symmetry. 
In the absence of a spin-orbit coupling, in this 
case we have a doublet with angular momentum and parity $J^p=1/2^+$, $3/2^+$. 

There is some preliminary evidence from the CLAS Collaboration for the 
existence of two peaks in the $nK^+$ invariant mass distribution at 1523 
and 1573 MeV \cite{battaglieri}. The mass difference between these two 
peaks is very close to the mass difference in the stringlike model between 
the ground state pentaquark at 1540 MeV (fitted) and the first excited 
state $\Theta_1$ at 1599 MeV. 

\subsection{Magnetic moments}

The magnetic moment of a multiquark system is given by the 
sum of the magnetic moments of its constituent parts 
\ba
\vec{\mu} \;=\; \vec{\mu}_{\rm spin} + \vec{\mu}_{\rm orb} \;=\; 
\sum_i \mu_i (2\vec{s}_{i} + \vec{\ell}_{i}) ~, 
\ea
where the quark magnetic moments $\mu_u$, $\mu_d$ and $\mu_s$ are determined 
from the proton, neutron and $\Lambda$ magnetic moments and 
satisfy $\mu_q=-\mu_{\bar{q}}$. 

The $SU_{\rm sf}(6)$ wave function of the ground state pentaquark 
has the general 
structure 
\ba
\psi_{A_2} \;=\; \left[ \psi^{\rm c}_{F_1} \times \left[ \psi^{\rm o}_{A_1} 
\times \psi^{\rm sf}_{F_2} \right]_{F_2} \right]_{A_2} ~. 
\ea
Since the ground state orbital wave function has $L^p=0^+$, the 
magnetic moment only depends on the spin part. For the $\Theta^+$ 
exotic state we obtain 
\ba
\mu_{\Theta^+} \;=\; (2\mu_u + 2\mu_d + \mu_s)/3 
\;=\; 0.382 \; \mu_N ~,
\label{mmneg}
\ea
in agreement with the result obtained \cite{Liu} for the MIT bag model. 
These results for the magnetic moments are independent of the 
orbital wave functions, and are valid for any quark model in which the 
eigenstates have good $SU_{\rm sf}(6)$ spin-flavor symmetry. 

\begin{table}
\centering
\caption{\small Magnetic moments of the ground state antidecuplet 
pentaquarks for $J^p=\frac{1}{2}^-$ in $\mu_N$.}
\label{mmpenta}
\vspace{15pt}
\begin{tabular}{lcr}
\hline
& & \\
& $\mu_{\rm th}$ & $\mu_{\rm calc}$ \\
& & \\
\hline
& & \\
$\Theta^+$ & $(6\mu_u + 6\mu_d + 3\mu_s)/9$ & $ 0.382$ \\
$N^0$      & $(5\mu_u + 6\mu_d + 4\mu_s)/9$ & $ 0.108$ \\
$N^+$      & $(6\mu_u + 5\mu_d + 4\mu_s)/9$ & $ 0.422$ \\
$\Sigma^-$ & $(4\mu_u + 6\mu_d + 5\mu_s)/9$ & $-0.166$ \\
$\Sigma^0$ & $(5\mu_u + 5\mu_d + 5\mu_s)/9$ & $ 0.148$ \\
$\Sigma^+$ & $(6\mu_u + 4\mu_d + 5\mu_s)/9$ & $ 0.462$ \\
$\Xi^{--}_{3/2}$ & $(3\mu_u + 6\mu_d + 6\mu_s)/9$ & $-0.440$ \\
$\Xi^{-}_{3/2}$  & $(4\mu_u + 5\mu_d + 6\mu_s)/9$ & $-0.126$ \\
$\Xi^{0}_{3/2}$  & $(5\mu_u + 4\mu_d + 6\mu_s)/9$ & $ 0.188$ \\
$\Xi^{+}_{3/2}$  & $(6\mu_u + 3\mu_d + 6\mu_s)/9$ & $ 0.502$ \\
& & \\
\hline
\end{tabular}
\end{table}

In Table~\ref{mmpenta}, we present the magnetic moments of all members of 
the ground state antidecuplet. The magnetic moments are typically an order 
of magnitude smaller than the proton magnetic moment. In addition, they 
satisfy the generalized Coleman-Glashow sum rules \cite{coleman,hong} 
for the antidecuplet 
\ba
\mu_{\Theta^+} + \mu_{\Xi^+_{3/2}} &=& \mu_{N^+} + \mu_{\Sigma^+} ~,
\nonumber\\
\mu_{\Theta^+} + \mu_{\Xi^{--}_{3/2}} &=& \mu_{N^0} + \mu_{\Sigma^-} ~, 
\nonumber\\
\mu_{\Xi^{--}_{3/2}} + \mu_{\Xi^+_{3/2}} &=& \mu_{\Xi^-_{3/2}} + 
\mu_{\Xi^0_{3/2}} ~,
\ea
and
\ba
2\mu_{\Sigma^0} \;=\; \mu_{\Sigma^-} + \mu_{\Sigma^+} 
\;=\; \mu_{N^0} + \mu_{\Xi^0_{3/2}} 
\;=\; \mu_{N^+} + \mu_{\Xi^-_{3/2}} ~. 
\ea
The same sum rules hold for the chiral quark-soliton model in the chiral 
limit \cite{Kim}. In the limit of equal quark masses $m_u=m_d=m_s=m$, the 
magnetic moments of the antidecuplet pentaquark states become proportional 
to the electric charges $\mu_i = Q_i / 18m$, which implies that the sum of 
the magnetic moments of all members of the antidecuplet vanishes 
$\sum_i \mu_i = 0$. 

\section{Summary and conclusions}

In this contribution, we have discussed a stringlike model of pentaquarks, 
in which the four quarks are located at the corners of an equilateral 
tetrahedron with the antiquark in its center. Geometrically this is the  
most stable configuration. The ground state pentaquark belongs 
to the flavor antidecuplet, has angular momentum and parity $J^p=1/2^-$ and, 
in comparison with the proton, has a small magnetic moment. The width is 
expected to be narrow due to a large suppression in the spatial overlap 
between the pentaquark and its decay products \cite{song}.

The first report of the discovery of the pentaquark has triggered an enormous 
amount of experimental and theoretical studies of the properties of exotic 
baryons. Nevertheless, there still exist many doubts and questions about the 
existence of this state, since in addition to various confirmations there are 
also several experiments in which no signal has been observed \cite{babar}. 
Hence, it is of the utmost importance to understand the origin between these 
apparently contradictory results, and to have irrefutable proof for or 
against its existence \cite{zhaoclose}.  
If confirmed, the measurement of the quantum numbers of the 
$\Theta(1540)$, especially the angular momentum and parity, and the excited 
pentaquark states, may help to distinguish between different models and to 
gain more insight into the relevant degrees of freedom and the underlying 
dynamics that determines the properties of exotic baryons.  

\section*{Acknowledgements}

This work is supported in part by a grant from CONACyT, M\'exico.

\end{document}